\documentclass[%
reprint,
superscriptaddress,
amsmath,amssymb,
aps,
]{revtex4-1}

\usepackage{color}
\usepackage{graphicx}
\usepackage{dcolumn}
\usepackage{bm}

\begin {document}

\title{Infinite invariant density in a semi-Markov process with continuous state variables}

\author{Takuma Akimoto}
\email{takuma@rs.tus.ac.jp}
\affiliation{%
  Department of Physics, Tokyo University of Science, Noda, Chiba 278-8510, Japan
}%

\author{Eli Barkai}
\affiliation{%
  Department of Physics, Bar-Ilan University, Ramat-Gan
}%

\author{G\"unter Radons}
\affiliation{%
  Institute of Physics, Chemnitz University of Technology, 09107 Chemnitz, Germany
}%


\date{\today}

\begin{abstract}
We report on a fundamental role of a non-normalized {\color{black}formal steady state,} i.e., an infinite invariant density, in a semi-Markov 
process 
where  the state  is determined by the inter-event time of successive {\color{black}renewals}. 
{\color{black}The state describes  certain  observables found in models of  anomalous diffusion, e.g.,  the velocity in the generalized 
L\'evy walk model and the energy of a particle in the trap model. In our model, the inter-event-time distribution follows a fat-tailed 
distribution, which makes the state value more likely to be zero because long inter-event times imply small state values. 
We find two scaling laws describing the density for the state value, which accumulates in the vicinity of zero in the long-time limit.  
These laws provide universal behaviors in the accumulation process and 
give the exact expression of the infinite invariant density.}
Moreover, we provide two distributional limit theorems for time-averaged observables in these non-stationary processes. {\color{black}
We show that the infinite invariant density plays an important role in determining the distribution of time averages.}
 
\end{abstract}

\maketitle


\section{Introduction}

There is a growing number of studies on applications of infinite invariant densities
in physical literature, 
ranging from deterministic dynamics describing intermittency \cite{Akimoto2007,Korabel2009, Akimoto2010, Akimoto2013b,Meyer2017}, models of laser cooling \cite{Bardou2002,Bardou1994,bertin2008,lutz2013}, anomalous diffusion \cite{Rebenshtok2014,Holz2015,Leibovich2019,wang2019ergodic,Vezzani2019}, {\color{black}fractal-time} 
renewal processes \cite{Wang2018}, 
and non-normalized Boltzmann states \cite{Aghion2019}. 
{\color{black}Infinite invariant densities are non-normalized formal steady states of systems and were 
   studied in dynamical systems exhibiting intermittency 
\cite{PM1979,*pomeau1980,*Manneville1980,Thaler1983, Aizawa1984,*Ai1984,*Aizawa1989,Thaler1995, Aaronson1997, Thaler2000, Zweimuller2004}.
The corresponding ergodic  theory 
is known as infinite ergodic theory, which is based on Markovian stochastic processes \cite{Darling1957, Lamperti1958}, and 
states that time averages of some observables do not converge to the corresponding ensemble averages but
 become random variables in the long-time limit 
\cite{Aaronson1981, Aaronson1997, Thaler1998, Thaler2002, TZ2006,Akimoto2008,Akimoto2015}. Thus, time averages cannot be replaced by 
ensemble averages even in the long-time limit. This striking feature is different from usual ergodic systems. Therefore,}  finding 
unexpected links between infinite ergodic theory and nonequilibrium phenomena attracts a significant interest in statistical 
physics \cite{Akimoto2007,Korabel2009,Akimoto2010, Korabel2012, *Korabel2013,Akimoto2010a,Akimoto2012, Rebenshtok2014,Holz2015,Leibovich2019, lutz2013, Aghion2019, Meyer2017a}.

{\color{black}In equilibrium systems, time averages of an observable converge to a constant, which is given by the ensemble 
average with respect to the  invariant probability measure, i.e., the equilibrium distribution. 
However, in nonequilibrium processes, this ergodic property sometimes does not hold.}
In particular,  distributional behaviors of time-averaged observables have been experimentally unveiled. Examples are the  
intensity of fluorescence in quantum dots \cite{Brok2003,stefani2009}, diffusion coefficients of a diffusing biomolecule  in living cells \cite{Golding2006,Weigel2011,Jeon2011,Manzo2015}, and interface fluctuations in Kardar-Parisi-Zhang universality class \cite{takeuchi2016}, where time averages of an observable, 
obtained from trajectories under the same experimental setup, do not converge to a constant but remain random.
These distributional behaviors of time averages of some observables have been investigated by several stochastic 
models describing anomalous diffusion processes \cite{God2001, He2008, Miyaguchi2011, *Miyaguchi2013, *Akimoto2013a, *Miyaguchi2015, Metzler2014,  Budini2016, *Budini2017,  AkimotoYamamoto2016, AkimotoYamamoto2016a, Akimoto2016,*Akimoto2018, Albers2014,Albers2016,Albers2018,Leibovich2019}. 
 
{\color{black}
While several works have considered applications of infinite ergodic theory to anomalous dynamics, 
one cannot apply infinite ergodic theory straightforwardly to stochastic processes. Therefore, 
our goal is to provide a deeper understanding of infinite ergodic theory in non-stationary stochastic processes. To this end, we 
 derive an exact form of the infinite invariant density and expose
 the role of the non-normalized steady state 
 in a minimal model for nonequilibrium non-stationary processes. 
 In particular, we unravel how the infinite invariant density plays a vital role in a semi-Markov process (SMP), which characterizes 
the velocity  of the generalized L\'evy walk (GLW) \cite{Shlesinger1987,Albers2018}. 

Our work  addresses three issues. Firstly, what is the propagator of the state variable?  In particular, we will show its relation to the mean number of renewals in the state variable.  
Secondly, we derive the exact form of the infinite invariant density, which is obtained from a formal steady state of the propagator 
found in the first part. Finally, we investigate distributional limit theorems of some time-averaged observables 
and discuss the role of the infinite invariant density. We end the paper with a summary. }

\section{Infinite ergodic theory in Brownian motion}

{\color{black}Before describing our stochastic model, we provide the infinite invariant density and its role in one of the simplest models of diffusion, i.e., 
Brownian motion.} 
Statistical properties of equilibrium systems {\color{black}or nonequilibrium systems with steady states}
are described by a normalized density describing the steady state.
On the other hand, a formal steady state sometimes cannot be normalized in nonequilibrium processes, {\color{black}where non-stationarity 
is essential} \cite{Bardou2002,Bardou1994,bertin2008, Rebenshtok2014,Holz2015,Leibovich2019,lutz2013, Aghion2019}. 
Let us consider a free 1D Brownian motion in infinite space. The formal steady state is a uniform distribution, which cannot be normalized in infinite space. {\color{black} To see this, consider the diffusion equation, 
$\partial_t P (x,t)= D \partial^2 _x P(x,t)$, where $P(x,t)$ is the density. Then, setting the left hand side to zero yields 
 a formal steady state, i.e., the uniform distribution. 
This is the simplest example of an infinite invariant density in nonequilibrium stochastic processes, where the system never reaches 
the equilibrium. 
Although the propagator of Brownian motion is known exactly, the role of the infinite invariant density is not so wel-known. 
Here, we will demonstrate its use. Later, we will see parallels and differences to the results for our SMP. }

First, we  consider the occupation time statistics. The 
classical arcsine law states that the ratio between the occupation time that a 1D Brownian particle spends on the positive side 
and the total measurement time   follows the arcsine distribution \cite{karatzas2012brownian}, {\color{black}which means that 
the ratio does not converge to a constant even in the long-time limit   and remains a random variable. 
Moreover, the ratio between the occupation time that a 1D Brownian particle spends on a region with a finite length and
the total measurement time does not converge to a constant. 
Instead, the normalized ratio exhibits intrinsic  trajectory-to-trajectory fluctuations 
and the distribution function follows a half-Gaussian, which is a special case of the Mittag-Leffler distribution known 
from the occupation time distribution for Markov chains \cite{Darling1957}. }

These two laws are distributional limit theorems 
for time-averaged observables because the occupation time can be represented by a sum of indicator functions. 
To see this, consider the Heaviside step function, i.e., $\theta(x) = 1$ if $x> 0$, otherwise zero. 
{\color{black}The occupation time on the positive side can be represented by $\int_0^t \theta (B_{t'})dt'$, where 
$B_t$ is a trajectory of a Brownian motion.
The integral of $\theta(x)$ with respect to the infinite invariant density, i.e., $\int_{-\infty}^\infty \theta (x)dx$, 
is clearly diverging. On the other hand, if we consider  
$f(x)=\theta (x - x_a )\theta (x_b -x )$, i.e.; it is one for $x_a < x < x_b$, and zero otherwise, the integral of $f(x)$ with respect to 
the infinite invariant density 
remains finite. Therefore, }
the observable for the arcsine law is not integrable with respect to the infinite density while that for {\color{black}the latter case} is integrable. 
Therefore,  the integrability of the observable  discriminates the two distributional limit theorems  in occupation time statistics.

{\color{black}For non-stationary processes, the propagator never reaches a steady state, i.e., equilibrium state. 
However, a formal steady state exists and 
is described by the infinite invariant density for many cases. This infinite invariant density will characterize 
distributional behaviors of time-averaged observables. 
For Brownian motion, this steady state is trivial (uniform) and in some sense non-interesting. 
However, we will show that this integrability condition is rather general as in infinite ergodic theory of dynamical systems.   
}

\section{semi-Markov process}

Here, we introduce a semi-Markov process (SMP), which {\color{black}couples a renewal process to an observable. A} renewal process is a point process where 
an inter-event time of two successive renewal points is an independent and identically distributed (IID) random variable \cite{Cox1962}. In SMPs,
 a state changes at renewal points. {\color{black}More precisely,} a state remains constant in between successive renewals. In what follows, we consider 
 continuous state variables. In particular, 
 the state is characterized by a continuous scalar variable and the scalar value is determined 
by the inter-event time. 
In this sense, the continuous-time random walk and a dichotomous process {\color{black} are SMPs \cite{metzler00, God2001}}. Moreover, time series of magnitudes/distances of earthquakes  
can be described by an SMP because 
there is a {\color{black}correlation} between the magnitude 
 and the inter-event time \cite{Helmstetter2002}. 
In the trap model \cite{bouchaud90} 
a random walker is trapped in random energy landscapes. Because  escape times from a trap are IID random variables depending on 
the trap and {\color{black}its} mean escape time is given by the energy {\color{black}depth of the trap}, 
the value of the energy depth is also described by an SMP.  
Therefore,  a state variable,  in a different context, can have {\color{black}many} meanings (see also Ref.~\cite{Meyer_2018}) 

As a typical physical example of this process, we {\color{black}consider} the GLW \cite{Shlesinger1987}. 
This system  can be applied to many physical systems such as turbulence dynamics and subrecoil laser cooling \cite{richardson1926,obukhov1959,Bardou2002,Bardou1994,bertin2008,Albers2018,aghion2018,Bothe2019}, where the state is considered to be velocity or momentum.  
In the GLW a walker moves with constant velocities $v_n$ over time segments of lengths $\tau_n$ between
turning points occurring at times $t_n$, i.e, $\tau_n = t_n - t_{n-1}$, where flight durations $\tau_n$ are IID random variables.
 Thus, the displacement $X_n$ in time segment $[t_{n-1}, t_n]$ is given by $X_n = v_n \tau_n$.  
 A coupling between $v_n$ and $\tau_n$ is given by joint probability density function (PDF) 
$\psi(v,\tau)$.
As a specific coupling which we consider in this paper, the absolute values $|v_n|$ of the velocities and flight durations $\tau_n$
in elementary flight events are coupled deterministically via 
\begin{equation}
|v_n| = \tau_n^{\nu -1}, 
\label{coupling-v-tau}
\end{equation}
or equivalently via 
\begin{equation}
\tau_n = |v_n|^{\frac{1}{\nu -1}}.
\end{equation} 
The quantity $\nu >0$ is an important parameter characterizing a given GLW. This nonlinear coupling was also considered in Ref.~\cite{Klaflater1987, Meyer2017a,aghion2018}.
The standard L\'evy walk corresponds to case $\nu=1$, implying that the velocity 
does not depend on the flight duration. In what follows, we focus on case $0<\nu<1$. {\color{black}Importantly, 
if $\tau \to \infty$, in this regime $|v_n| \to 0$. Thus, we will find accumulation of density in the vicinity of $v=0$. 
This is because we assume a power-law distribution for flight durations, that favors long flight durations. }
Some investigations such as Refs.~\cite{Shlesinger1987,Albers2018} concentrated on the behavior in
coordinate space, where a trajectory $x(t)$ is a piecewise linear function
of time $t$. 

In the following, we  {\color{black}denote the state variable as  velocity 
 and investigate the velocity distribution at time $t$,} where a trajectory of velocity $v(t)$ is a piecewise constant
function of $t$. 
An SMP consists of a sequence $\left\{ \mathcal{E}_{1},%
\mathcal{E}_{2},\ldots \right\} $ of elementary flight events $\mathcal{E}%
_{n}=(v_n, \tau_n)$. We note that this sequence $\mathcal{E}_{n}$ ($n=1, \cdots$) is an IID random {\color{black}vector} variable.
Thus, the velocity process of a GLW is characterized by the
joint PDF of velocity $v$ and flight duration $\tau$ in an elementary flight event: 
\begin{equation}
{\color{black}\phi}  (v,\tau )=\left\langle \delta \left( v-v_{i}\right) \delta \left( \tau
-\tau_{i}\right) \right\rangle  \label{jpdgen}
\end{equation}%
The symbol $\delta \left(.\right) $ denotes the Dirac delta function. 
PDF $\psi (\tau )$ of the flight durations is defined through the marginal density of the joint PDF $\psi (v,\tau )$:
\begin{equation}
\psi (\tau )=\int_{-\infty }^{+\infty } {\color{black}\phi}  (v,\tau) dv=\left\langle \delta \left( \tau -\tau_{i}\right) \right\rangle .
\label{psi}
\end{equation}%
Similarly one can get PDF ${\color{black}\chi} (v)$ for the velocities
of an elementary event as%
\begin{equation}
{\color{black}\chi}  (v)=\int_{0}^{+\infty } {\color{black}\phi}  (v,\tau )\;d\tau =\left\langle
\delta \left( v-v_{i}\right) \right\rangle .  \label{chi}
\end{equation}%

In L\'{e}vy walk treatments usually $\psi (\tau )$ is prescribed and
chosen as a slowly decaying function with a power-law tail: 
\begin{equation}
\psi (\tau )  \sim \frac{c}{|\Gamma(-\gamma)|}  \tau ^{-1-\gamma }\quad (\tau\to\infty)  
\label{psigamma}
\end{equation}%
with the parameter $\gamma >0$ characterizing the algebraic decay and $c$ being a scale parameter. 
{\color{black}A pair of  parameters $\nu$ and $\gamma$  determines} the essential
properties of the GLW and the asymptotic behavior in the velocity space. 
Of special interest is the regime $0<\gamma <1$. There the sequence of renewal points $\left\{t_{n},n=0,1,2,\ldots \right\} $, at which  velocity $v(t)$ changes, i.e., 
\begin{equation}
t_{n}=\sum_{i=1}^{n}\tau_{i}  
\label{renewal points}
\end{equation}
with $t_0=0$,  is a non-stationary process in the sense that the rate of change is not constant but varies with time \cite{God2001,Bardou2002}. This is because
 the mean flight duration diverges, i.e., $\left\langle \tau_{i}\right\rangle
=\int_{0}^{\infty }\tau \;\psi (\tau )\;d\tau =\infty$. 
To determine the last velocity $v(t)$ at time $t$, one needs  to know the time interval straddling $t$, which is defined as  $\tau \equiv  t_{n+1} -t_n$ with $t_n < t<t_{n+1}$ and was discussed in \cite{Barkai2014,AkimotoYamamoto2016a}. 
{\color{black}In other words, to determine the distribution of the velocity at time $t$,
one needs to know the time when the first renewal occurs after time $t$ and the time for the last renewal event before $t$.}


\section{General expression for the propagator}

\subsection{Standard derivation}

We are interested in the propagator $p(v,t)$, which is the PDF
of finding a velocity $v$ at time $t$, given that the process started at $t=0$ with $v=0$. 
 To derive an expression for $p(v,t)$, we note that
at every renewal time $t_{n-1}$ the process starts anew with velocity $v_n$ 
until $t<t_n$. So one needs the {\color{black}probability $R(t)dt$ of 
finding some renewal event in $[t, t+dt)$.} This quantity is called sprinkling density in the
literature \cite{Bardou2002} and {\color{black}it is} closely related to the
renewal density in renewal theory \cite{Cox1962}. 

It is obtained from a recursion relation for the PDF
 $R_{n}(t)=\left\langle \delta \left( t-t_{n}\right) \right\rangle $
that the $n$-th renewal point $t_{n}$ occurs exactly at time $t$. Using the PDF $\psi(\tau)$, we get the iteration rule
\begin{equation}
R_{n+1}(t)=\int_{0}^{t}dt^{\prime }\;\psi (t-t^{\prime
})R_{n}(t^{\prime })  \label{recursion}
\end{equation}%
with the initial condition $R_{0}(t)=\delta (t)$, which means that we assume a renewal occurs at $t=0$, i.e., 
{\it ordinary renewal process} \cite{Cox1962}. 
Summing both sides from $n=0$ to infinity, one gets the equation  of $R(t)\equiv
\sum_{n=0}^{\infty }R_{n}(t)$ for $t>0$, i.e., 
\begin{equation}
R(t)=\int_{0}^{t}dt^{\prime }\;\psi (t-t^{\prime })R(t^{\prime
})+ R_0(t).  
\label{renewal equation}
\end{equation}%
Eq.~(\ref{renewal equation}) is known as the renewal
equation. The solution of this equation is easily obtained in Laplace space as 
\begin{equation}
\widetilde{R}(s)=\frac{1}{1-\widetilde{\psi }(s)},
\label{Laplace renewal density}
\end{equation}%
where $\widetilde{R}(s)=\int_{0}^{\infty }R(t)\exp (-st)\;dt$. 
The integral of $R(t)$ is related to the expected number of renewal events $\left\langle N(t)\right\rangle $
occurring up to time $t$, i.e., 
\begin{equation}
\left\langle N(t)\right\rangle =\int_{0}^{t}R(t^{\prime
})\;dt^{\prime }.  \label{mean number of renewals}
\end{equation}%
Note that here the event at $t=0$ is also counted while the event at $t=0$ is often {\color{black}excluded}  in renewal
theory. 

With knowledge of $R(t) $, which in principle can be obtained by Laplace inversion of Eq.~(\ref{Laplace renewal density}), one can formulate the solution of the propagator as 
\begin{equation}
p(v,t)=\int_{0}^{t}dt^{\prime }W(v,t-t^{\prime })R(t^{\prime }),
\label{propagator}
\end{equation}%
where $W(v,t-t^{\prime })$ takes into account the last incompleted  flight
event, starting at the last renewal time $t^{\prime }$, provided that the flight duration is longer than
 $t-t'$ with velocity $v$. 
Thus, $W(v,t)$ is given by 
\begin{eqnarray}
W(v,t) \equiv \int_{t}^{\infty }d\tau {\color{black}\phi} (v,\tau) .
\label{constrained sojourn}
\end{eqnarray}%
Integrating this over all velocities leads to the survival probability $\Psi (t)$ 
of the sojourn time, i.e., the probability that an event lasts longer than a given time $t$ 
\begin{equation}
\Psi (t)=\int_{-\infty }^{+\infty
}W(v,t)\;dv=\int_{t}^{\infty }d\tau \;\psi (\tau).  
\label{sojourn}
\end{equation}%
Using {\color{black} Eqs.~(\ref{chi}),} (\ref{Laplace renewal density}), and (\ref{constrained sojourn}) 
one can write down the propagator in Laplace space
\begin{equation}
\widetilde{p}(v,s)=\widetilde{W}(v,s)\widetilde{R}(s)=\frac{1}{s}\frac{{\color{black}\chi}(v)-\widetilde{{\color{black}\phi} }(v,s)}{1-\widetilde{\psi }(s)}.  
\label{Montroll Weiss}
\end{equation}%
This is a general expression of the propagator 
and an analogue of the Montroll-Weiss equation of the continuous-time random walk \cite{Montroll1965}. Recalling 
 $\int dv \left[ {\color{black}\chi} (v)-\widetilde{{\color{black}\phi} }(v,s)\right] =1-\widetilde{\psi }(s)$ gives $\int dv\;\widetilde{p}(v,s)
 =1/s$,
implying that propagator $p(v,t)$ in the form of Eq.~(\ref{Montroll Weiss}) is
correctly normalized $\int dv\;p(v,t)=1$.

In what follows, as a specific example, we consider a deterministic coupling between $\tau_i$ and $|v_i|$. 
The joint PDF ${\color{black}\phi} (v,\tau )$ is specified as follows: flight duration $\tau_{i}$
is chosen randomly from the PDF $\psi (\tau )$, and the corresponding
absolute value of the velocity $\left\vert v_{i}\right\vert $ is 
deterministically given by $\left\vert v_{i}\right\vert =$ $\tau_{i}^{\nu -1}$. 
Finally, the sign of $v_{i}$ is determined with equal probability, implying that 
\begin{equation}
{\color{black}\phi} (v,\tau )=\frac{1}{2}\left[ \delta \left( v-\tau ^{\nu -1}\right) +\delta \left(
v+\tau ^{\nu -1}\right) \right] \;\psi (\tau )  \label{jpd1}
\end{equation}
with ${\color{black}\phi} (v,\tau )={\color{black}\phi} (-v,\tau )$. Alternatively, one can specify the velocity 
first using the PDF ${\color{black}\chi} (v)= {\color{black}\chi} (-v)$. Then, one can express the joint PDF
 ${\color{black}\phi} (v,\tau )$ also as 
\begin{equation}
{\color{black}\phi} (v,\tau )=\delta \left( \tau -\left\vert v\right\vert ^{\frac{1}{\nu -1}}\right) {\color{black}\chi} (v).  
\label{jpd2}
\end{equation}%
Although Eqs.~(\ref{jpd1}) and (\ref{jpd2}) are equivalent, the latter
suggests a different interpretation {\color{black}of selecting an elementary event, e.g. 
 the velocity is selected from $\chi(v)$ first 
and then this velocity state lasts for duration $\tau_{i}=\left\vert v_{i}\right\vert ^{\frac{1}{\nu -1}}$.}
Obviously prescribing $\psi (\tau )$ determines ${\color{black}\chi} (v)$ [via Eqs.(\ref{chi}) and (\ref{jpd1})] 
and vice versa [via Eqs.(\ref{psi}) and (\ref{jpd2})]. From Eq.~(\ref{constrained sojourn}), one gets 
\begin{eqnarray}
W(v,t) ={\color{black}\chi} (v) \theta (|v|^{\frac{1}{\nu -1}}-t),
\label{constrained sojourn2}
\end{eqnarray}%
 {\color{black}where $\theta(x)$ is the Heaviside step function. }

Before deriving our main results, we give the equilibrium distribution of the propagator for $\gamma>1$.   
Although we assumed $\gamma<1$ in Eq.~(\ref{psigamma}), the general expression for the propagator, Eq.~(\ref{Montroll Weiss}), 
is exact also for $\gamma >1$. 
For $\gamma > 1$, the mean flight duration $\langle \tau \rangle$ is finite and we have
\begin{equation}
p_{\rm eq}(v) = \lim_{s\to0} s \widetilde{p}(v,s) = \frac{\int_0^\infty {\color{black}\phi} (v,\tau)\tau d\tau}{\langle \tau \rangle}.
\end{equation}
Therefore, for $\gamma>1$, as expected, the equilibrium distribution exists; i.e., the propagator reaches a steady state:
\begin{equation}
p(v,t) \to  p_{\rm eq}(v) = \frac{{\color{black}\chi}(v)|v|^{\frac{1}{\nu-1}}}{\langle \tau \rangle} 
\label{steady-state}
\end{equation}
for $t\to\infty$. Here, we note that the equilibrium distribution has a different form {\color{black}for the decoupled case,}
 i.e., ${\color{black}\phi}(v,\tau) = {\color{black}\chi}(v) \psi(\tau)$. 
 In this case, it is easily obtained as $p_{\rm eq}(v) ={\color{black}\chi}(v)$.

Because the integration of $R(t)$ gives $\langle N(t) \rangle$, we get an exact expression 
for the propagator 
\begin{equation}
p(v,t)={\color{black}\chi} (v)\;\left[ \left\langle N(t)\right\rangle -\left\langle
N(t-t_{c}(v))\right\rangle\right] ,  
\label{pvt}
\end{equation}%
where $t_c(v)\equiv |v|^{\frac{1}{\nu -1}}$. We note that $\left\langle N(t)\right\rangle=0$ when $t<0$.  
In particular, one can express $p(v,t)$ as
\begin{equation}
p(v,t)= {\color{black}\chi} (v)\cdot \left\{ 
\begin{tabular}{l}
$\left\langle N(t)\right\rangle $ \\ 
\\
$\left\langle N(t)\right\rangle -\left\langle N(t-t_{c}(v))\right\rangle $%
\end{tabular}%
\right. 
\begin{tabular}{l}
for \\ 
\\
for%
\end{tabular}%
\begin{tabular}{l}
$t<t_{c}(v)$ \\ 
\\
$t>t_{c}(v)$%
\end{tabular}%
. \label{pvt_piecewise}
\end{equation}%
Since  we have {\color{black}made} no approximation, the solution is {\color{black}formally exact, while 
the remaining  difficulty is to obtain $\left\langle N(t)\right\rangle$. }
This is the central result of this section. 

The mean number $\left\langle
N(t)\right\rangle $ of renewals up to time $t$ increases monotonically from $%
\left\langle N(t\rightarrow 0)\right\rangle =1$ because the first jump is at 
$t_{0}=0+$, which implies that $\lim_{t\rightarrow 0}p(v,t)={\color{black}\chi} (v)$, which
is the velocity distribution of the elementary event as given by Eq.~(\ref{chi}). 
For a given velocity $v$ satisfying $t<t_{c}(v)$, 
 the function $p(v,t)$ increases until $t$ reaches $t_{c}(v)$  because 
$\left\langle N(t)\right\rangle $ is a monotonically increasing function.
  Thereafter $p(v,t)$ stays
constant or decreases because $\left\langle N(t)\right\rangle -\left\langle
N(t-t_{c}(v))\right\rangle $ stays constant or decreases depending on
whether the renewal sequences $\left\{ t_{n},n=0,1,2,\ldots \right\} $ are
equilibrium sequences or not \cite{Cox1962,Akimoto2014,Miyaguchi2016,Akimoto2018a}. This in turn depends on the shape of $\psi
(\tau )$, more precisely on the decay of $\psi (\tau )$ for large $\tau $,
as detailed below. 

For a discussion of the velocity profile $p(v,t)$ for a
fixed time $t$, it is more convenient to rewrite Eq.~(\ref{pvt}) for $v>0$ as 
\begin{equation}
p(v,t)={\color{black}\chi} (v)\;\left[ \left\langle N(t)\right\rangle -\left\langle
N(t-t_{c}(v))\right\rangle \theta (v-v_{c}(t))\right]  
\label{pvt_decr}
\end{equation}%
where we introduced the critical velocity $v_{c}(t)=t^{\nu -1}$ and 
 $v_{c}(t)$ is monotonically {\color{black}decreasing} as function of $t$ because $0<\nu<1$. 
 For negative $v$, $p(v,t)$ follows from the symmetry $p(v,t)=p(-v,t)$.
Thus, for a fixed $t$ and  $\left\vert v\right\vert <v_{c}(t)$, 
$p(v,t)$ is the same as ${\color{black}\chi} (v),$ enlarged by the velocity-independent
factor $\left\langle N(t)\right\rangle $, whereas for $\left\vert
v\right\vert >v_{c}(t)$ it has a non-trivial $v$-dependence \ due to the $v$-dependence 
of $\left\langle N(t-t_{c}(v))\right\rangle$.  Note that at
velocity $v=v_{c}(t)$ the profile of $v$ jumps by the value 
\begin{equation}
\delta p=\lim_{\varepsilon
\rightarrow 0}[p(v_{c}(t)-\varepsilon ,t)-p(v_{c}(t)+\varepsilon ,t)]= {\color{black}\chi}(v_{c}(t))
\label{jump}
\end{equation}
at the critical velocity $v=v_{c}(t)$ because we assume  $\left\langle
N(0)\right\rangle =1$.

\subsection{Another derivation of Eq.~(\ref{pvt})}

Here, we give another derivation of the propagator, i.e., Eq.~(\ref{pvt}). 
{\color{black}The joint PDF of $v(t)$ with $t$ satisfying   $t_n <t < t_{n+1}$, denoted by $p_n (v,t)$, }
can be written as 
\begin{equation}
p_n (v,t) = \langle \delta ( \tau_{n+1} - |v|^{\frac{1}{\nu-1}}) I(t_n <t<t_{n+1}) \rangle,
\end{equation}
where $I(\cdot ) =1$ if the condition in the bracket is satisfied, and 0 otherwise.  It follows that the propagator can be obtained as 
{\color{black}a sum over the number of renewals $n$:}
\begin{equation}
p(v,t ) = \sum_{n=0}^\infty p_n (v,t) .
\end{equation}
Using ${\color{black}\chi}(v)$ and $t_{n+1}=t_n + \tau_{n+1}$, we have 
\begin{equation}
p(v,t )= {\color{black}\chi}(v) \sum_{n=0}^\infty \langle  I(t_n <t<t_{n}+|v|^{\frac{1}{\nu-1}}) \rangle, 
\end{equation}
where we note that $\langle I(t_n <t ) \rangle$ gives a probability:
\begin{equation}
\langle I(t_n <t ) \rangle = \Pr (N(t) > {\color{black}n} ).
\end{equation}
The mean of $N(t)$ can be written as 
\begin{equation}
\langle N(t) \rangle = \sum_{n=1}^\infty n \Pr (N(t) = n) = \sum_{{\color{black}n=0}}^\infty  \Pr (N(t) > n),
\end{equation}
where we used identity $\Pr (N(t) = n) =  \Pr (N(t) > n-1) -  \Pr (N(t) > n)$. 
Therefore, we have Eq.~(\ref{pvt}).

\begin{figure}
\includegraphics[width=.9\linewidth, angle=0]{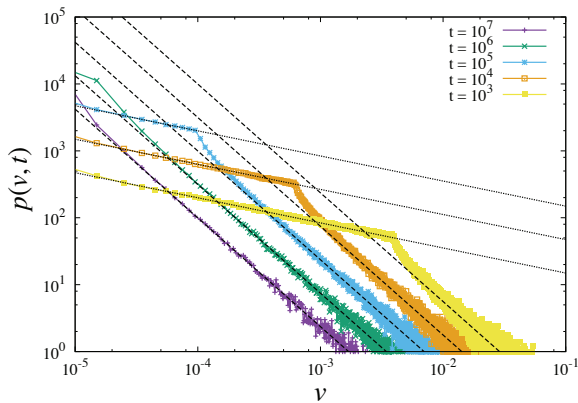}
\caption{Time evolution of the propagator for different times ($\gamma =0.5$ and $\nu=0.2$). Symbols with lines are the results of numerical 
simulations. Dashed and dotted lines are the theories, i.e., Eqs.~(\ref{pv_continuous}) and (\ref{propagator_asympt1}), respectively. Infinite density 
can be observed for $v >  v_c(t)$ while the propagator follows a different scaling, i.e., Eq.~(\ref{propagator_asympt1}), for $v <  v_c(t)$. 
We used the PDF $\psi(\tau)=\gamma \tau^{-1-\gamma}$ for $\tau\geq 1$ {\color{black}and $\psi(\tau)=0$ for $\tau<1$} as the {\color{black}flight-duration} PDF.
 }
\label{propagator}
\end{figure}

\section{Infinite invariant density in a semi-Markov process}
\subsection{Infinite invariant density}
To proceed with the discussion of Eq.~(\ref{pvt_decr}), we use Eq.~(\ref{psigamma}) 
for the {\color{black}flight-duration} PDF and consider $\gamma<1$. The PDF of velocities ${\color{black}\chi} (v)$ in an elementary event
can be obtained by Eqs.~(\ref{chi}) and (\ref{jpd1}): 
\begin{equation}
{\color{black}\chi} (v)=\frac{1}{2}\psi (\left\vert v\right\vert ^{\frac{1}{\nu -1}})\frac{\left\vert v\right\vert ^{-1-\frac{1}{1-\nu }}}{1-\nu  }
.  \label{chigen}
\end{equation}%
For the specific choice for $\psi (\tau )$ given in Eq.~(\ref{psigamma})  
the  asymptotic form with  $0<\nu <1$ yields 
\begin{equation}
{\color{black}\chi} (v)\sim \frac{c}{2 ( 1-\nu )  |\Gamma(-\gamma)|}
\left\vert v\right\vert ^{-1+\frac{\gamma }{1-\nu }}\text{ for }v\rightarrow 0  .
\label{chizero}
\end{equation}%

First, we give the asymptotic behavior of $\langle N(t) \rangle$ for $t\to\infty$. 
Because the Laplace transform of $\psi(\tau)$ is given by $\widetilde{\psi }(s) = 1- c s^{\gamma } + o(s^\gamma)$ 
for  $s\to 0$,  Eqs.~(\ref{Laplace renewal density}) and (\ref{mean number of renewals}) yields {\color{black}the well-known result:}
\begin{equation}
\left\langle N(t)\right\rangle \sim \frac{1}{c\Gamma(1+\gamma) } 
t^{\gamma }~~\text{ for }t\rightarrow \infty.  \label{asymptotic mean}
\end{equation}
{\color{black}However, for our purposes, we need to go beyond this limit as shown below.}  
The renewal function gives the exact form of the propagator [see Eq.~(\ref{pvt})]. 
There are two regimes in the propagator as seen in Eq.~(\ref{pvt_piecewise}). For $t<t_c(v)$, or equivalently $v<v_c(t)$,  
the propagator is given by 
\begin{equation}
p(v,t)= \langle N(t) \rangle {\color{black}\chi} (v) .
\end{equation}
In this regime the propagator is an increasing function of $t$ because $\langle N(t) \rangle$ is a monotonically increasing function  
whose asymptotic behavior is given by Eq.~(\ref{asymptotic mean}), whereas the support $(-v_{c}(t), v_{c}(t))$ will shrink because 
$v_c(t)=t^{-(1-\nu)} \to 0$ as $t\to\infty$. {\color{black}For $t\gg 1$ and $v <  v_c(t)$, implying $v <  v_c(t) \ll 1$}, the propagator becomes
\begin{equation}
p(v,t)  \sim \frac{\sin (\pi \gamma) t^{\gamma }}{2\pi (1-\nu )  }  |v|^{-1+\frac{\gamma }{1-\nu }}.
\label{propagator_asympt1}
\end{equation}
This is a universal law in the sense that the asymptotic form does not depend on the detailed form of the {\color{black}flight-duration} PDF
such as  scale parameter $c$. 

For $t>t_c(v)$, or equivalently $v>v_c(t)$,  the propagator is given through 
 $\left\langle N(t)\right\rangle -\left\langle N(t-t_{c}(v))\right\rangle $. {\color{black}For $t\gg 1$ and $v_c(v)<v \ll1$,} 
 \begin{eqnarray}
\left\langle N(t)\right\rangle -\left\langle N(t-t_{c}(v))\right\rangle &\cong& \frac{t^{\gamma} - (t- t_{c}(v))^\gamma }{c\Gamma(\gamma +1) } 
\label{Nt-Nt1}\\
&\sim& \frac{t^{\gamma -1} t_{c}(v)}{c\Gamma(\gamma) } .
\end{eqnarray}
 Therefore, the asymptotic behavior of the propagator becomes 
\begin{equation}
p(v,t)\sim {\color{black}\chi} (v)\left\vert v\right\vert ^{\frac{1}{\nu -1}}\frac{t^{\gamma -1}}{c\Gamma(\gamma)  }~~\text{ for }t\rightarrow \infty,
  \label{pv_continuous}
\end{equation}%
which can also be obtained simply by changing $\langle \tau \rangle$ in Eq.~(\ref{steady-state}) into $\int_0^t \tau \psi (\tau)d\tau$ 
except for the proportional constant. Here, 
 one can define a formal steady state ${\color{black} I_\infty(v)}$ using Eq.~(\ref{pv_continuous}) as follows:
\begin{equation}
{\color{black} I_\infty(v)} \equiv  \lim_{t\to \infty} t^{1-\gamma} p(v,t) = \frac{ {\color{black}\chi}(v) \left\vert v\right\vert ^{\frac{1}{\nu -1}}}{c\Gamma(\gamma) }.
\label{inf-d}
\end{equation}
This does not depend on $t$ in the long-time limit and {\color{black}is} a natural extension of the steady state for $\gamma>1$, i.e., Eq.~(\ref{steady-state}). 
{\color{black}In this sense, $I_{\infty}(x)$ is a formal steady state of the system. 
However, ${\color{black} I_\infty(v)} $ is not normalizable and thus it is sometimes called  {\it infinite 
invariant density}.} 
Using Eq.~(\ref{chizero}),  the asymptotic form of the infinite density for $v\ll 1$ becomes 
\begin{equation}
{\color{black} I_\infty(v)}  \sim \frac{\gamma \sin (\pi \gamma )}{2\pi (1-\nu) } |v|^{-1 -\frac{1-\gamma}{1-\nu}}.
\label{ifd_asympt}
\end{equation}
We note that the infinite density describes the propagator only for $v>v_c(t)$. While $v_c(t) \to 0$ in the long-time limit, 
the propagator for $v\ll 1$ is composed of two parts, i.e., Eqs.~(\ref{propagator_asympt1}) and {\color{black}(\ref{pv_continuous}).  
These behaviors are illustrated in Fig.~\ref{propagator}, where the support of the propagator is restricted to $|v|<1$. 
In particular, the accumulation at zero velocity for $v<v_c(t)$ and a trace of the infinite density for $v>v_c(t)$ are clearly shown.}
In general, the propagator for $v\gg 1$ is described by the small-$\tau$ behavior of the {\color{black}flight-duration} PDF through Eq.~(\ref{pvt_piecewise}). 

\begin{widetext}
\subsection{Scaling function}
Rescaling $v$ by $v'=t^{1-\nu} v$ in the propagator, we find a scaling function. In particular, the rescaled propagator 
does not depend on time $t$ and approaches the scaling function denoted by $\rho(v')$ in the long-time limit ($t\to\infty$): 
\begin{equation}
{\color{black} p_{\rm res} (v',t) \equiv p(v'/t^{1-\nu},t) \left| \frac{dv}{dv'}\right|}
\to \rho (v') \equiv \left\{
\begin{array}{ll}
 \dfrac{ \sin (\pi \gamma )}{2\pi (1-\nu ) } |v'|^{-1+\frac{\gamma }{1-\nu }}~&(v'<1)\\
 \\
 \dfrac{\sin (\pi \gamma ) \{ 1 - (1-v'^{\frac{1}{\color{black}\nu-1}})^\gamma\}}{2\pi (1-\nu)} |v'|^{-1 + \frac{\gamma}{1-\nu}} &(v' \geq 1) ,
 \end{array}
 \right.
 \label{master-curve}
\end{equation}
where we used Eq.~(\ref{Nt-Nt1}) and note that $v_c'\equiv t^{1-\nu}v_c(t)=1$. {\color{black}In the scaling function, the long-time 
limit is taken in advance. Thus, the scaling function describes only small-$v$ behaviors of $p(v,t)$. In other words, large-$v$ 
behaviors of $\rho(v)$ are not matched with those of $I_\infty(v)$ while large-$v$ behaviors of $\rho(v)$ are matched 
with small-$v$ behaviors of $I_\infty(v)$. 
The scaling function is normalized and}
continuous at $v'=1$ whereas
$p(v,t)$ is not continuous at $v=v_c(t)$ {\color{black}for finite $t$ 
because the jump in the propagator at $v=v_c$ is given by Eq.~(\ref{jump}) and $\chi(v_c(t)) \to 0$ for $t\to \infty$.} 
As shown in Fig.~\ref{propagator-rescale}, 
 rescaled propagators at different times $t$ coincide with the scaling function for $t\gg 1$. 
{\color{black}Note that the scaling function describes the behavior of $p(v,t)$ for $v \ll 1$ in the long-time limit, which
 does not capture the behavior of $p(v,t)$ for $v > 1$. Large-$v$ behaviors of  $p(v,t)$ can be 
 described by $I_\infty (v)$. Although $I_\infty (v)$ 
  depends on the details of $\chi(v)$, the scaling function 
is not sensitive to all the details except for $\gamma$. In this sense, it is a general result. 

\begin{figure*}
\includegraphics[width=.95\linewidth, angle=0]{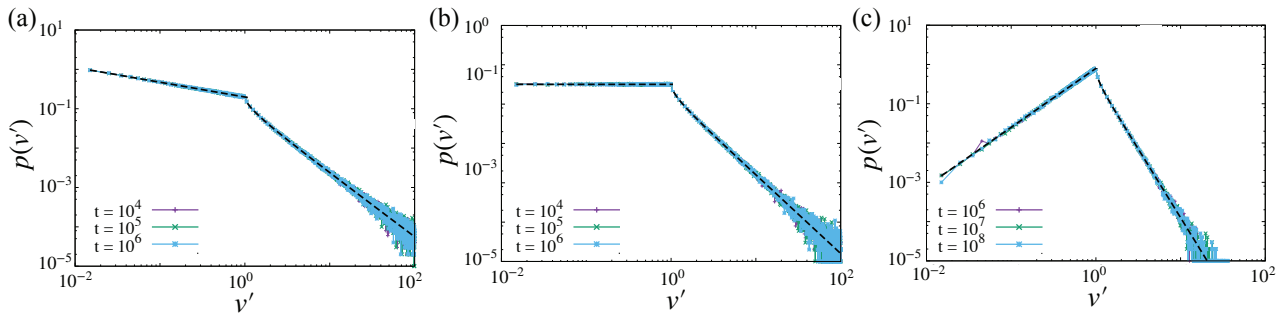}
\caption{Rescaled propagators  for (a) $\nu=0.2$, (b) $\nu=0.5$ and (c) $\nu=0.8$ ($\gamma =0.5$). 
Symbols with lines are the results of numerical 
simulations for different times $t$. Dashed lines represent the scaling functions, i.e., Eq.~(\ref{master-curve}). 
{\color{black}The {\color{black}flight-duration} PDF is the same as that in Fig.~\ref{propagator}.}}
\label{propagator-rescale}
\end{figure*}

\subsection{Ensemble averages}

The theory of infinite ergodic theory is a theory of observables. This means that we must classify different 
observables and define the limiting laws with which their respective ensemble averages are obtained
in the long time limit. 
We will soon consider also time averages. Consider the observable $f(v)$. 
The corresponding ensemble average is given by 
\begin{equation}
\langle f(v(t)) \rangle \equiv \int_{-\infty}^\infty f(v) p(v,t)dv
= \int_{-v_c(t)}^{v_c(t)} p(v,t) f(v) dv + \int_{v_c(t)}^\infty p(v,t) f(v)dv
+  \int_{-\infty}^{-v_c(t)} p(v,t) f(v) dv.
\label{ensemble-ave}
\end{equation}
If we take the time $t\to\infty$, we have 
\begin{equation}
\langle f(v(t)) \rangle 
\cong \int_{-1}^{1} \rho(v') f(v'/t^{1-\nu}) dv' + t^{\gamma-1} \int_{v_c(t)}^\infty I_\infty (v) f(v) dv
+ t^{\gamma-1} \int_{-\infty}^{-v_c(t)} I_\infty (v) f(v) dv ,
\end{equation}
where we performed a change of variable and used the scaling function in the first term, and we also used 
$p(v,t) \cong t^{\gamma-1} I_\infty (v)$ for $|v|>v_c(t)$ in the second and third terms. 
Moreover, we assume that the second and third term in Eq.~(\ref{ensemble-ave}) does not diverge. In what follows, we consider
$f(v) = |v|^\alpha$.  When $f(v)$ is integrable with respect to $\rho(v)$, i.e., 
$\int_{-\infty}^\infty \rho(v) f(v) dv<\infty$, $\alpha$ satisfies the following inequality: 
\begin{equation}
-\frac{\gamma}{1-\nu} < \alpha < \frac{1-\gamma}{1-\nu}. 
\end{equation}
In this case, the leading term of the asymptotic behavior of the ensemble average is given by the first term: 
\begin{equation}
\langle f(v(t)) \rangle \sim t^{-\alpha (1-\nu)} \int_{-1}^1 \rho(v) f(v) dv \quad (t \to \infty), 
\label{en-ave-scaling}
\end{equation}
where we used Eq.~(\ref{master-curve}):
\begin{equation}
\int_{-1}^{1} \rho(v) f(v/t^{1-\nu})dv \sim t^{-\alpha (1-\nu)} \int_{-1}^1 \rho(v) |v|^\alpha dv \quad (t \to \infty).
\end{equation}
Thus, the ensemble average goes to zero and infinity in the long-time limit for $\alpha >0$ and $\alpha <0$, respectively. 
On the other hand, 
when $f(v)$ is integrable with respect to $I_\infty (v)$, i.e., $\int_{-\infty}^\infty I_\infty (v) f(v) dv < \infty$,  
where  $f(v)$ satisfies $f(v) \sim v^\alpha$ with $\alpha > \frac{1-\gamma}{1-\nu}>0$ for $v\to 0$, 
the second and third terms becomes 
\begin{equation}
 t^{\gamma-1} \int_{v_c(t)}^\infty I_\infty (v) f(v) dv+ t^{\gamma-1} \int_{-\infty}^{-v_c(t)} I_\infty (v) f(v) dv
 =  t^{\gamma-1} \int_{-\infty}^\infty I_\infty (v) f(v) dv,
\end{equation}
Because the relation between $\alpha, \nu$ and $\gamma$ satisfies 
$\alpha (1-\nu) > 1-\gamma$, the asymptotic behavior of the ensemble average is given by
\begin{equation}
 \langle f(v(t)) \rangle \sim  t^{\gamma-1} \int_{-\infty}^\infty I_\infty (v) f(v) dv \quad (t \to \infty).
 \label{en-ave-infty}
\end{equation}
A structure of Eqs.~(\ref{en-ave-scaling}) and (\ref{en-ave-infty}) is very similar to an ordinary  equilibrium averaging  in the sense that 
there is a time-independent average with respect to $\rho(v)$ or $I_\infty (v)$ on the right hand side, where the choice of 
$\rho(v)$ or $I_\infty (v)$ depends on whether the observable is integrable with respect to $\rho(v)$ or $I_\infty (v)$. 
The beauty of infinite ergodic theory is that this can be extended to time averages, which as mentioned will be discussed below. 

In the long-time limit, $p(v,t)$ behaves like a delta distribution in the following sense:
\begin{equation}
\int_{-\infty}^\infty \lim_{t\to\infty} p(v,t)f(v) dv = \int_{-\infty}^\infty \rho(x) \lim_{t\to\infty}  f(x/t^{1-\nu}) dx =f(0).
\label{delta-function}
\end{equation}
Eq.~(\ref{delta-function}) is clearly obtained when $f(v)$ is integrable with respect to $\rho(v)$, i.e., 
$\int_{-\infty}^\infty \rho(v) f(v)dv < \infty$. Even when $f(v)$ is not integrable with respect to $\rho(v)$, 
Eq.~(\ref{delta-function}) is valid if $f(v)$ is integrable with respect to $I_\infty (v)$. 
In fact, the asymptotic behavior of the ensemble average $\langle f(v(t)) \rangle$ becomes 
$\langle f(v(t)) \rangle \to 0 =f(0)$ for $t\to \infty$, as shown  above. 
 Therefore, Eq.~(\ref{delta-function}) is valid in this case. When both integrals diverge, 
Eq.~(\ref{delta-function}) is no longer valid. However, if there exists a positive constant $\varepsilon$ such that $\psi (\tau)=0$ for $\tau < \varepsilon$, 
Eq.~(\ref{delta-function}) is always valid. 
In the long-time limit, the ensemble average is trivial in the sense that it simply gives the value of the observable at $v=0$. 
At this stage, there is no replacement of a ``steady state" concept. However, 
in general, scaling function $\rho (v)$ describes the propagator near $v=0$ while 
infinite invariant density $I_\infty (v)$ describes  the propagator for $v>0$ including large-$v$ behaviors. 
Therefore, as shown in Eqs.~(\ref{en-ave-scaling}) and (\ref{en-ave-infty}), 
both the scaling function and the infinite invariant density play an important role for the evaluation of 
certain ensemble averages at time $t$.
}

\end{widetext}

\section{Distributional limit theorems}

When the system is stationary, a time average {\color{black}approaches} a constant in the long-time limit, 
which implies ergodicity of the system.
However, time averages of some observables may not converge to a constant but {\color{black}properly scaled
time averages converge} in distribution when the system is non-stationary {\color{black}as it is case for $\gamma<1$.} 
While we focus on  regime $0 < \nu <1$, the following theorems can be extended to  regime $\nu>1$. 

\begin{figure}
\includegraphics[width=.9\linewidth, angle=0]{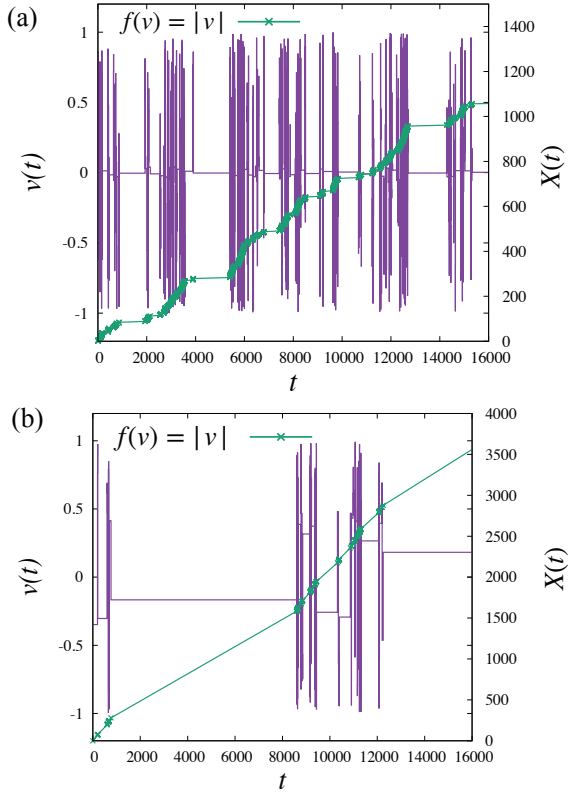}
\caption{Trajectories of  velocity $v(t)$ and the integral of a function $f(v)=|v|$, i.e., $X(t)=\int_0^t |v(t')| dt'$. 
Because  velocity $v(t)$ is a piece-wise constant function, integral $X(t)$ is a piece-wise linear function of $t$.
Parameter sets $(\gamma, \nu)$ are $(0.8, 0.2)$  and $(0.5, 0.8)$ for (a) and (b), respectively.
 }
\label{trajectory}
\end{figure}

To obtain the distribution of these time averages, we consider the propagator of 
the {\color{black}integrals} of these observables along a trajectory from $0$ to $t$, denoted by $X(t)$, which 
are piece-wise-linear functions of $t$ and 
can be described by 
a {\it continuous accumulation process} (see Fig.~\ref{trajectory}) \cite{Akimoto2015}. 
Time average of function $f(v)$ is defined by
\begin{equation}
\overline {f}(t) \equiv \frac{1}{t} \int_0^t f(v (t')) dt' = \frac{X(t)}{t}.
\end{equation}
{\color{black}As specific examples, we will consider time averages of the absolute value of the velocity and the squared velocity, i.e., 
 $f(v) = |v|$ or $f(v) =v^2$. Integrated value }
$X(t)$ can be represented by
\begin{equation}
X(t)=\sum_{n=1}^{{\color{black}N(t)-1}} f(v_n) \tau_n + f(v_{{\color{black}N(t)}})(t - t_{{\color{black}N(t)-1}}).
\end{equation} 
{\color{black}The stochastic process of $X(t)$} can be characterized by  a recursion relation, which is the same as in the derivation of the velocity propagator. 
Let $R_f(x,t)$ be the PDF of $x=X(t)$ when a renewal occurs exactly at time $t$, then we have
\begin{equation}
R_f(x,t) =  \int_0^x dx' \int_0^t dt' {\color{black}\phi}_{f} (x', t') R_f(x-x', t-t') + R_f^0(x,t), 
\end{equation}
where ${\color{black}\phi}_{f} (x,\tau )= \delta \left( x- f(\tau ^{\nu-1 })\tau \right) \psi (\tau )$ and $R_f^0(x,t)=\delta(x)\delta(t)$. 
Here, we assume that function $f(v)$ is an even function. 
We note that we use a deterministic coupling between $\tau$ and $v$, i.e., Eq.~(\ref{coupling-v-tau}).
The PDF of $X(t)$ at time $t$ is given by
\begin{eqnarray}
P_f(x,t) &=&  \int_0^x dx' \int_0^t  dt' {\color{black}\Phi}_{f} (x', t') R_f(x-x', t-t'),  
\end{eqnarray}
where
\begin{equation}
{\color{black}\Phi}_{f} (x,t) = \int_t^\infty d\tau  \psi (\tau) \delta(x- f(\tau^{\nu -1}) t).
\end{equation}

The double-Laplace transform with respect to $x$ and $t$ yields
\begin{equation}
\widetilde{P}_f(k,s) =  \frac{\widetilde{{\color{black}\Phi}}_{f}(k,s)}{1- \widetilde{{\color{black}\phi}}_{f}(k,s)}, 
\label{montroll-weiss-like}
\end{equation}
where $\widetilde{{\color{black}\phi}}_{f}(k,s)$ and $\widetilde{{\color{black}\Phi}}_{f}(k,s)$ are 
the double-Laplace transforms of ${\color{black}\phi}_{f} (x,\tau )$ and ${\color{black}\Phi}_{f} (x,t)$  given by 
\begin{equation}
\widetilde{\color{black}\phi}_{f}(k,s) = \int_0^\infty d\tau e^{-s\tau -kf(\tau ^{\nu-1 })\tau} \psi(\tau)
\label{psi_laplace_ta}
\end{equation}
and
\begin{equation}
\widetilde{{\color{black}\Phi}}_{f}(k,s) = \int_0^\infty dt e^{-st } \int_t^\infty d\tau  e^{-kf(\tau ^{\nu-1 })t} \psi (\tau),
\label{PSI_laplace_ta}
\end{equation}
respectively. Eq.~(\ref{montroll-weiss-like}) is the exact form of the PDF of $X(t)$ in Laplace space.  

\begin{widetext}
Before considering a specific form of  $f(v)$, we show that there are two different classes of distributional limit theorems 
of time averages. 
Expanding $e^{-k f(\tau^{\nu-1})\tau}$ in Eq.~(\ref{psi_laplace_ta}), we have
\begin{equation}
\widetilde{{\color{black}\phi}}_{f}(k,s) \cong \widetilde{ \psi} (s) -  k \int_0^\infty d\tau f(\tau^{\nu-1})\tau \psi(\tau) e^{-s\tau}+ O(k^2).
\end{equation}
Using Eq.~(\ref{chigen}), one can write the second term with {\color{black}$s\to0$} as
\begin{equation}
\int_0^\infty d\tau f(\tau^{\nu-1})\tau \psi(\tau) = \frac{1}{1-\nu} \int_0^\infty dv  f(v) v^{\frac{3-\nu}{\nu-1}} \psi(v^{\frac{1}{\nu-1}})
=2 \int_0^\infty f(v) v^{\frac{1}{\nu-1}} {\color{black}\chi} (v)dv = 2 c\Gamma(\gamma) \int_0^\infty f(v) {\color{black}I_\infty} (v)dv.
\end{equation}
When  $f(v)$ is integrable with respect to the infinite invariant density, i.e., $\int_0^\infty f(v) I_\infty(v) dv < \infty$, 
the second term is still finite for $s\to 0$. As shown below, we will see that the integrability gives a condition that determines the shape of the 
distribution function for the normalized time average, i.e.,  $\overline{f}(t)/\langle \overline{f}(t) \rangle$. 
\end{widetext}

\subsection{Time average of the absolute value of $v$}

In this section, we show that there are two phases for distributional behaviors of time averages. The phase line is determined by  a relation between $\gamma$ and $\nu$. 
As a specific  choice of function $f(v)$, we consider the absolute value of the velocity, i.e., $f(v)=|v|$. Thus, $X(t)$ is given by
\begin{equation}
X(t) = \sum_{n=1}^{{\color{black}N(t)-1}} \tau_n^{\nu } + \tau_{{\color{black}N(t)}}^{\nu -1}  (t - t_{{\color{black}N(t)-1}}),
\end{equation} 

For $\nu < \gamma$, the moment $\langle \tau^\nu \rangle$ is finite, i.e., $\langle \tau^\nu \rangle  < \infty$.
This condition is equivalent to the following condition represented by the infinite density:
\begin{equation}
\langle f(v) \rangle_{\rm inf} = \int_0^\infty f(v) {\color{black}I_\infty (v)} dv < \infty. 
\label{inf-condition}
\end{equation}
The double Laplace transform $\widetilde{P}_{|v|}(k,s) $ is calculated in Appendix~C (see Eq.~(\ref{montroll-weiss-like1})).  For $s \to 0$, the leading term 
of $ -\left. \frac{\partial \widetilde{P}_{|v|}(k,s)}{\partial k}  \right|_{k=0} $ becomes 
\begin{equation}
-\left. \frac{\partial \widetilde{P}_{|v|}(k,s)}{\partial k}  \right|_{k=0} 
\sim \frac{\langle \tau^\nu \rangle}{ c s^{1+\gamma}}.
\end{equation}
It follows that the mean of $X(t)$ for $t\to \infty$ becomes
\begin{equation}
\langle X(t) \rangle \sim \frac{\langle \tau^\nu \rangle}{c\Gamma (1+\gamma)} t^\gamma.
\label{mean-tav}
\end{equation}
Since the mean of $X(t)$ increases with $t^\gamma$, we consider a situation where $k\sim s^\gamma$ for small 
$k,s \ll 1$ in the double-Laplace space. Thus, all the term $k/s^\nu$ ($\ll 1$) and $O(k^2/s^\gamma)$ in Eq.~(\ref{montroll-weiss-like1}) 
can be ignored. It follows that the asymptotic form of $\widetilde{P}(k,s)$ is given by
\begin{equation}
\widetilde{P}_{|v|}(k,s) = \frac{ c s^{\gamma -1} /\langle \tau^\nu \rangle}{ k + c s^\gamma /\langle \tau^\nu \rangle }.
\end{equation}
 This is the  double Laplace transform of PDF $G_t'(\langle \tau^\nu \rangle x/c)$ \cite{Feller1971}, where 
\begin{equation}
G_t (x) = 1 - L_\gamma (t/x^{1/\gamma})
\end{equation}
and $L_\gamma (x)$ is a one sided L\'evy distribution; i.e., the Laplace transform of PDF $l_\gamma (x) \equiv L_\gamma'(x)$ is given by $e^{-k^\gamma}$. 
By a straightforward calculation one obtain the asymptotic behavior of the second moment as follows:
\begin{equation}
\langle X(t)^2 \rangle \sim \frac{2 \langle \tau^{\nu} \rangle^2 t^{2\gamma}}{c^2 \Gamma (1+ 2\gamma)} .
\label{second-tav}
\end{equation}
Furthermore, the $n$th moment can be represented by 
\begin{equation}
\langle X(t)^n \rangle \sim \frac{n! \Gamma (1+\gamma)^n}{\Gamma (1+ n\gamma)} \langle X(t) \rangle^n
\end{equation}
for $t\to\infty$. 
It follows that random variable $X(t) / \langle X(t) \rangle$ converges in distribution to a random variable 
$M_\gamma$ whose PDF follows the {\it Mittag-Leffler distribution} {\color{black}of} order $\gamma$, where
\begin{equation}
\langle e^{-z M_\gamma} \rangle \sim \sum_{n=0}^\infty \frac{ \Gamma (1+\gamma)^n}{\Gamma (1+ n\gamma)} (-z)^n.
\end{equation}
In other words, the normalized time averages defined by $\langle \tau^\nu \rangle X(t)/(c t^\gamma)$ do not converge to a constant but 
the PDF converge to a non-trivial  distribution (the {\it Mittag-Leffler distribution}).  In particular, 
the PDF can be represented through the L\'evy distribution:
\begin{equation}
G_1 '(x) = \frac{1}{\gamma}x^{-\frac{1}{\gamma}-1} l_\gamma (x^{-1/\gamma})
\end{equation}
To quantify trajectory-to-trajectory fluctuations of the time averages, we consider the ergodicity breaking (EB) parameter \cite{He2008} defined by 
\begin{equation}
{\rm EB}(t) \equiv \frac{ \langle \overline{f}(t)^2 \rangle - \langle \overline{f}(t) \rangle^2}{\langle \overline{f}(t) \rangle^2 } ,
\end{equation}
where $\langle \cdot \rangle$ implies the average with respect to the initial condition. 
When the system is ergodic, it  goes to zero as $t\to\infty$. On the other hand, it converges to a non-zero constant when 
the trajectory-to-trajectory fluctuations are intrinsic. For $\nu < \gamma <1$, the EB parameter becomes
\begin{equation}
{\rm EB}(t) \to {\rm ML}(\gamma) \equiv \frac{2\Gamma(1+\gamma)^2}{\Gamma (1+2\gamma)} -1 \quad (t\to \infty),
\label{eb-ML}
\end{equation}
which means that time averages do not converge to a constant but {\color{black}they become} 
a random variable with a non-zero variance. 
For $\gamma>1$, the EB parameter actually goes to zero in the long-time limit. Moreover, it also goes to zero as $\gamma \to 1$ 
in Eq.~(\ref{eb-ML}). 
We note that the condition (\ref{inf-condition}) is general in a sense that the distribution of time averages of function $f(v)$ satisfying the 
condition (\ref{inf-condition}) follows the Mittag-Leffler distribution, {\color{black}which is the same condition as in infinite ergodic theory 
\cite{Aaronson1997}.}

For $\nu > \gamma$, $\langle \tau^\nu \rangle$ diverges and equivalently $\langle f(v) \rangle_{\rm inf}=\infty$, 
which results in a distinct behavior of the time averages. 
 Using Eq.~(\ref{coeff-mean}), we have 
\begin{equation}
-\left. \frac{\partial \widetilde{P}_{|v|}(k,s)}{\partial k}  \right|_{k=0} 
\sim \frac{\gamma\Gamma (\nu -\gamma)}{(1+\gamma -\nu)\Gamma(1-\gamma)} \frac{1}{s^{1+\nu}}
\end{equation}
for $s\to 0$. The inverse Laplace transform gives  
\begin{equation}
\langle X(t) \rangle 
\sim \frac{\gamma |\Gamma (\nu -\gamma-1)|}{\Gamma(1-\gamma)\Gamma (1+\nu)} t^\nu
\label{moment_nu}
\end{equation}
for $t \to \infty$. 
Therefore, $X(t)$ scales as $t^\nu$, which means that  all the terms of $k/s^\nu$ in Eq.~(\ref{montroll-weiss-like2}) cannot be ignored.
These terms give the higher order moments. Performing the inverse Laplace transform of terms proportional to $1/s^{1+\nu}$ 
gives 
\begin{equation}
\langle X(t)^n \rangle \propto t^{n\nu}
\label{moments_largernu}
\end{equation}
for $t\to \infty$. 
By Eq.~(\ref{M2_nu}),  the EB parameter becomes 
\begin{widetext}
\begin{equation}
{\rm EB}(t) \to A(\gamma,\nu)\equiv \frac{2 (1+\gamma -\nu)\Gamma(1+\nu)^2}{\Gamma(1+2\nu)} 
\left[\frac{ (1+\gamma -\nu)\Gamma (2\nu -\gamma) \Gamma (1-\gamma) }{\gamma (2-2\nu + \gamma)  \Gamma(\nu -\gamma)^2}
+ 1 \right] -1 \quad (t\to \infty).
\label{eb-abs-inf}
\end{equation}
\end{widetext}
This EB parameter depends on $\gamma$ as well as $\nu$ ($>\gamma$) {\color{black}and was found also in Ref.~\cite{Albers2016}. }
{\color{black}We note that $A(\gamma,\nu)$ is a decreasing function of $\nu$. 
Therefore, trajectory-to-trajectory fluctuations of the time averages becomes insignificant for large $\nu$. 
In particular, $A(\gamma,\nu)$ converges to ${\rm ML}(\gamma)$ and 0 for $\nu \to \gamma +0$ and $\nu\to 1-0$, respectively. 
 In other words, the system becomes ergodic in the sense that the time averages converge to a constant 
 in the limit of $\gamma \to 1$ (and $\nu \to 1$). 
}

\subsection{Time average of the squared velocity}
For $f(v)= v^2$, $X(t)$ can be represented by
\begin{equation}
X(t) = \sum_{n=1}^{{\color{black}N(t)-1}} \tau_n^{2\nu-1 } + \tau_{{\color{black}N(t)} }^{2\nu -2} (t - t_{{\color{black}N(t)-1}}).
\end{equation} 
By the same calculation as in the previous case, using ${\color{black}\phi}_{v^2} (z,\tau )= \delta \left( z-\tau ^{2\nu-1 }\right) \psi (\tau ) $
and ${\color{black}\Phi}_{v^2} (z,t) = \int_t^\infty d\tau  \psi (\tau) \delta(z-\tau^{2\nu -2} t)$, one can 
express the double Laplace transform of $P(z,t)$ as 
\begin{equation}
\widetilde{P}_{v^2}(k,s) =  \frac{\widetilde{{\color{black}\Phi}}_{v^2}(k,s)}{1- \widetilde{\psi}_{v^2}(k,s)}. 
\end{equation}
Therefore, the limit distribution of $X(t)/\langle X(t) \rangle$ can be obtained in the same way as for the previous observable. 
In particular, the Mittag-Leffler distribution is a universal distribution of the normalized 
time average of $v^2$ if $2\nu -1 < \gamma$, i.e., $f(v)=v^2$ is integrable with respect to the infinite invariant density. 
On the other hand, the distribution of normalized time averages $X(t)/\langle X(t) \rangle$ becomes  
another distribution for $t\to\infty$  if $2\nu -1 > \gamma$ (see Appendix.~C).
It follows that $\langle X(t) \rangle \propto t^{2\nu -1}$ for $t\to\infty$ and 
the EB parameter becomes
\begin{equation}
{\rm EB}(t) \to  A(\gamma, 2\nu -1).
\label{eb-v2-inf}
\end{equation}
{\color{black}This expression was also obtained in Ref.~\cite{Albers2016}.}
 The exponent $2\nu-1$ in Eq.~(\ref{eb-v2-inf}) is different from that found in the EB parameter for $f(v)=|v|$ with $\nu > \gamma$. 
{\color{black}Therefore, our distributional limit theorem is not universal but depends on the observable.} 
On the other hand, the exponent $\gamma$ in the EB parameter for $2\nu -1 < \gamma$ is the same as that for $f(v)=|v|$ with 
 $\nu < \gamma$. 
 
\begin{figure}
\includegraphics[width=.9\linewidth, angle=0]{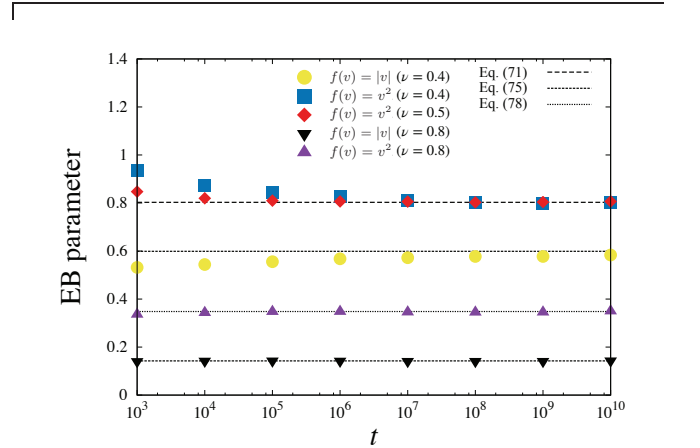}
\caption{Ergodicity breaking parameter as a function of the measurement time for  $\nu=0.4$, 0.5, and $\nu=0.8$ ($\gamma=0.3$). 
We note that $\nu=0.4$ and 0.5 satisfy $2\nu -1 <\gamma$ while $\nu=0.8$ satisfies $2\nu -1 > \gamma$. 
Symbols represent the results of numerical simulations. The long-dashed line represents Eq.~(\ref{eb-ML}), the 
two dotted lines represent Eq.~(\ref{eb-abs-inf}), and the dotted line represent Eq.~(\ref{eb-v2-inf}). 
{\color{black}The {\color{black}flight-duration} PDF is the same as that in Fig.~\ref{propagator}.}
 }
\label{eb}
\end{figure}

Figure~\ref{eb} shows that our theory works very well for both observables. 
For $f(v)=v^2$ with $\nu=0.4$ and $\nu=0.5$ ($\gamma=0.3$), both of which satisfy $2\nu -1 < \gamma$, 
the EB parameters do not depend on $\nu$. Moreover, Fig.~\ref{eb} shows that {\color{black}
the EB parameter given by  $A(\gamma, \nu)$ 
 is a decreasing function of $\nu$ for $\gamma<\nu$. 
}
  
\begin{figure*}
\includegraphics[width=.9\linewidth, angle=0]{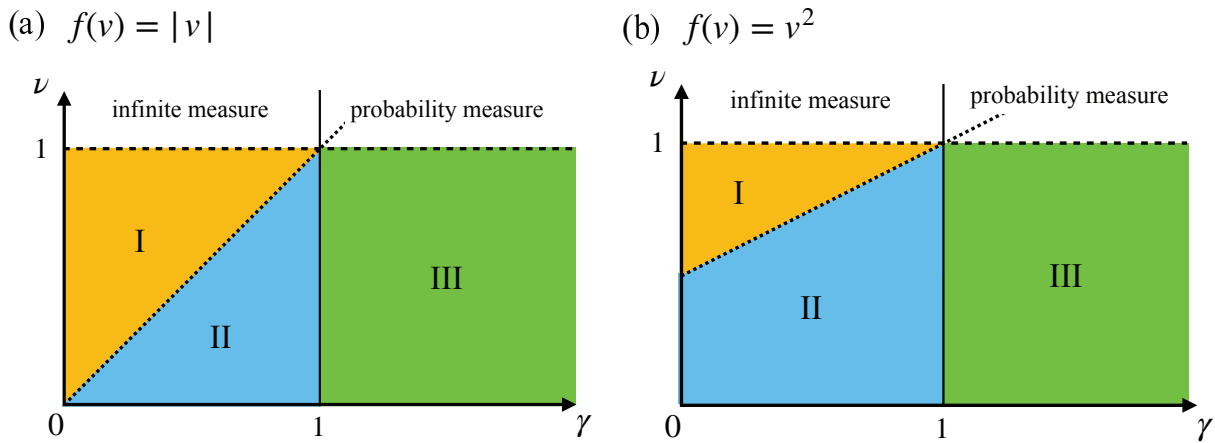}
\caption{Phase diagram of the parameter space $(\gamma,\nu)$ for (a)  $f(v)=|v|$ and (b)  $f(v)=v^2$. The solid line $\gamma=1$ 
describes the boundary of the infinite measure. The dotted line represents 
the boundary that the average of the observable $f(v)$ with respect to the infinite/probability measure diverges. 
For $\int f(v) p_{\rm eq} (v) dv<\infty$
 and $\gamma>1$ (region III), the time average converges to a constant, implying the EB parameter goes to zero. 
For $\int f(v) {\color{black} I_\infty(v)}  dv<\infty$ and $\gamma<1$ (region II),  the EB parameter becomes a non-zero constant given by Eq.~(\ref{eb-ML}), implying the time averages remain random variables. 
For $\int f(v) {\color{black} I_\infty(v)}  dv=\infty$ and $\gamma<1$ (region I),  the EB parameter becomes a non-zero constant given by Eq.~(\ref{eb-abs-inf}) or (\ref{eb-v2-inf}), implying the time averages remain random variables and it depends on $\gamma$ as well as $\nu$, which is different from case $\int f(v) {\color{black} I_\infty(v)} <\infty$ and $\gamma<1$. 
}
\label{phase}
\end{figure*}
 
\section{conclusion}
We investigated the propagator {\color{black} in an SMP and
 provided its exact form, which is described by  the mean number of renewals [see Eq.~(\ref{pvt_decr})]. 
 We assumed that $\chi(v)=\chi(-v)$ and that this function has support on zero velocity. More specifically, the 
 relation $v= \tau^{\nu-1}$ implies that long flight durations favor velocity close to zero since $0<\nu<1$ and 
 this is the reason for an accumulation of probability in the vicinity of zero velocity in this model. We prove that }
 the propagator accumulates in the vicinity of zero velocity in the long-time limit 
 when the mean {\color{black}flight-duration} diverges ($\gamma<1$) and the coupling parameter fulfills $\nu <1$.  
Taking a closer look at the vicinity of $v=0$,  
we found {\color{black}universal behaviors in the  asymptotic forms of the propagator. 
In particular the asymptotic behavior of the propagator for $v\ll 1$ follows two scaling laws, i.e., the infinite invariant density 
Eq.~(\ref{inf-d}) and the scaling function Eq.~(\ref{master-curve}). The scaling function describes a detailed 
structure of the propagator near $v=0$ including zero velocity 
while the infinite invariant density describes the propagator outside $v_c=t^{\nu-1}$.
Clearly $v_c \to 0$ when $t\to \infty$, and interestingly the asymptotic form outside $v_c$ 
becomes a universal form that is unbounded at the origin and cannot be normalized, i.e., an 
{\it infinite invariant density}. One advantage of considering the topic with an SMP is that we can attain an explicit expression 
for the infinite invariant density Eq.~(\ref{inf-d}). In contrast in general it is hard to find exact infinite invariant measures in 
deterministic dynamical systems, for example in the context of the Pomeau-Manniville map \cite{PM1979,*pomeau1980,*Manneville1980}.

Further, while the Mittag-Leffler distribution describing the distribution of time averages of 
 integrable observables is well known, from the Aaronson-Darling-Kac theorem, we considered here also
another distributional limit theorem [see Eqs.~(\ref{eb-abs-inf}) and (\ref{eb-v2-inf})]  which describes the distribution of time averages of 
certain non-integrable observables. 
Therefore, the integrability of the observable with respect to the infinite invariant density establishes a criterion on the type 
of distributional limit law, which is similar to findings in infinite ergodic theory. 
These results will pave the way for constructing  physics of non-stationary processes.}
Finally, we summarize our results by the phase diagram shown in Fig.~\ref{phase}. 
The infinite invariant density is always observed for $\gamma <1$. On the other hand, the boundary of the regions I and II depends on the 
observation function $f(v)$. 


\section*{Acknowledgement} 
This work was supported by JSPS KAKENHI Grant Number 16KT0021, 18K03468 (TA). 
EB thanks the Israel Science Foundation and Humboldt Foundation for support.

\appendix

\begin{figure}
\includegraphics[width=.9\linewidth, angle=0]{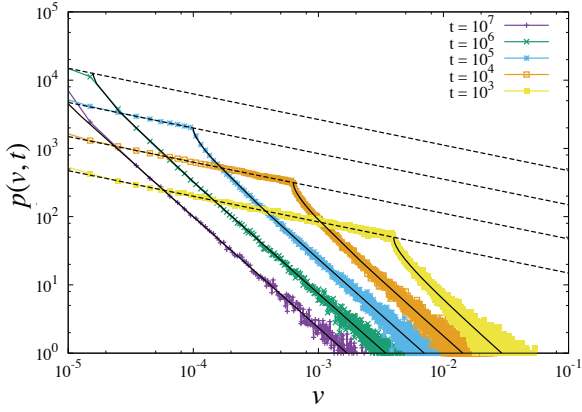}
\caption{Time evolution of the propagator for different times ($\gamma =0.5$ and $\nu=0.2$). Symbols with lines are the results of numerical 
simulations, which is the same as those in Fig.~\ref{propagator}. 
Dashed and solid lines are the theories, i.e., Eqs.~(\ref{pv_continuous}) and (\ref{prop-ML}), respectively. 
{\color{black}The {\color{black}flight-duration} PDF is the same as that in Fig.~\ref{propagator}.}
 }
\label{propagator-ML}
\end{figure}

\section{Exact form of the propagator outside $[-v_c(t), v_c(t)]$}

Here, we consider a specific form for the {\color{black}flight-duration} PDF to obtain the exact form of 
the propagator outside $[-v_c(t), v_c(t)]$. 
As a specific form, we use 
\begin{equation}
\psi (\tau) = -\frac{d}{dt} E_\gamma(-t^\gamma)= \frac{1}{\tau^{1-\gamma}} \sum_{n=0}^\infty 
 \frac{(-1)^n\tau^{n\gamma}}{\Gamma(\gamma n + \gamma)}, 
\label{flight-PDF-ML}
\end{equation}
where $E_\gamma(z) $ is the Mittag-Leffler function with parameter $\gamma$ defined as \cite{gorenflo2008continuous} 
\begin{equation}
E_\gamma(z) \equiv \sum_{n=0}^\infty \frac{z^n}{\Gamma (\gamma n+1)}. 
\end{equation}
In fact, the asymptotic behavior is given by a power law  \cite{gorenflo2008continuous}, i.e., 
\begin{equation}
\psi (\tau) \sim \frac{\Gamma(\gamma +1) \sin (\gamma \pi)}{\pi} \tau^{-1-\gamma}\quad (\tau\to\infty).
\end{equation}
Moreover, it is known that the Laplace transform of $\psi(\tau)$ is given by 
\begin{equation}
\widetilde{\psi}(s) = \frac{1}{1+s^\gamma}. 
\end{equation}
Therefore, the Laplace transform of $\langle N(t) \rangle$ becomes
\begin{equation}
\frac{1}{s(1-\widetilde{\psi}(s))} = \frac{1}{s^{1+\gamma}} + \frac{1}{s},
\end{equation}
and its inverse Laplace transform yields 
\begin{equation}
\langle N(t) \rangle = \frac{1}{\Gamma(1+\gamma)}t^\gamma + 1
\end{equation}
for any $t>0$. For $v\ll 1$, ${\color{black}\chi}(v)$ is given by
\begin{equation}
{\color{black}\chi} (v)\sim \frac{1}{2 ( 1-\nu )  |\Gamma(-\gamma)|}
\left\vert v\right\vert ^{-1+\frac{\gamma }{1-\nu }} .
\label{chizero-ML}
\end{equation}%
It follows that the propagator outside  $[-v_c(t), v_c(t)]$ becomes
\begin{equation}
p (v, t) \sim \frac{t^\gamma - (t-v^{\frac{1}{\nu-1}})^\gamma}{2 ( 1-\nu ) \sin (\gamma \pi)} \pi |v|^{-1 +\frac{\gamma}{1-\nu}}.
\label{prop-ML}
\end{equation}%
for $t\gg 1$ and $v> t^{\nu-1}$. As shown in Fig.~\ref{propagator-ML}, the propagator outside $[-v_c(t), v_c(t)]$ 
is described by Eq.~(\ref{prop-ML}), whereas we did not use Eq.~(\ref{flight-PDF-ML}). 


\begin{widetext}

\section{another proof of the asymptotic behavior of the propagator of $v$}

To obtain the propagator, i.e., the PDF of velocity $v$ at time $t$, it is almost equivalent to have the PDF $\psi_t (\tau)$ of  time interval straddling $t$, i.e., 
$\tau_{{\color{black}N(t)-1}}$, where  ${\color{black}N(t)-1}$ is the number of renewals until $t$ (not counting the one at $t_0=0$). 
In ordinary renewal processes, the double Laplace transform of the PDF with respect to $\tau$ and $t$ is given by \cite{Barkai2014}
\begin{equation}
\widetilde{{\color{black}\phi}} (k,s) = \frac{ \widetilde{\psi}(k) - \widetilde{\psi}(k+s)}{s [ 1 - \widetilde{\psi}(s)]}.
\end{equation}
For $\gamma <1$, the asymptotic behavior of this inverse Laplace transform can be calculated using a technique from Ref.~\cite{God2001}. 
For $t$ and $\tau \gg 1$, 
\begin{equation}
\psi_t (\tau) \sim \left\{
\begin{array}{ll}
\dfrac{\sin \pi \gamma}{\pi} \dfrac{t^\gamma}{{\color{black}\tau}^{1+\gamma}} \left[ 1 - \left(1-\dfrac{{\color{black}\tau}}{t} \right)^\gamma \right]\quad &({\color{black}\tau}<t)\\
\\
\dfrac{\sin \pi \gamma}{\pi} \dfrac{t^\gamma}{{\color{black}\tau}^{1+\gamma}}  &({\color{black}\tau}>t).
\end{array}
\right.
\label{waiting-time-t}
\end{equation}
This is the asymptotic result, which does not depend on the details of the {\color{black}flight-duration} 
PDF, i.e. different {\color{black}flight-duration} PDFs give the same result if the power-law 
exponent $\gamma$ is the same. On the other hand, detail forms of  $\psi_t (\tau)$, e.g., the behavior for small $t$ and $\tau$, 
depend on details of the {\color{black}flight-duration} PDF \cite{Wang2018}. 

Here, we consider a situation where the relation between the velocity and the flight duration is given by $|v|=\tau^{\nu -1}$. 
The PDF of velocity $v$ at time $t$, i.e., the propagator, can be represented through the PDF $\psi_t (\tau)$:
\begin{equation}
p (v,t) = \frac{1}{2|\nu -1|} |v|^{\frac{1}{\nu -1}-1} \psi_t (|v|^{\frac{1}{\nu -1}}).
\end{equation}
Note that $p (v,t)$ is symmetric with respect to $v=0$. Using Eq.~(\ref{waiting-time-t}) yields
\begin{equation}
p (v,t) \sim \left\{
\begin{array}{ll}
\dfrac{\sin \pi \gamma}{2\pi |1-\nu|} t^\gamma |v|^{-1+ \frac{\gamma}{1-\nu}} \left[ 1 - \left(1-\dfrac{|v|^{\frac{1}{\nu -1}}}{t} \right)^\gamma \right]\quad &(|v|>t^{\nu -1})\\
\\
\dfrac{  \sin \pi \gamma }{2\pi |1-\nu|} t^\gamma |v|^{-1+ \frac{\gamma}{1-\nu}}  &(|v| < t^{\nu-1}).
\end{array}
\right.
\end{equation}
The asymptotic form for $\nu<1$ becomes 
\begin{equation}
p (v,t) \sim \left\{
\begin{array}{ll}
\dfrac{  \sin \pi \gamma }{2\pi |1-\nu|} t^{\gamma -1}|v|^{-1+ \frac{\gamma}{1-\nu}}  &(|v| \ll  t^{\nu-1})\\
\\
\dfrac{\gamma \sin \pi \gamma}{2\pi (\nu -1)} t^{\gamma } |v|^{-1+ \frac{1-\gamma}{\nu-1}} & (t^{\nu-1} \ll |v|)
\end{array}
\right.
\end{equation}
for $t\to\infty$. Therefore, this is consistent with the propagator we obtained in this paper, Eqs.~(\ref{propagator_asympt1}) and (\ref{pv_continuous}). 
 
 For $\gamma >1$, the PDF $\psi_t (\tau)$ has an equilibrium distribution, i.e., for $t\to\infty$ the PDF $\psi_t(\tau)$ is given by
 \begin{equation}
 \psi_t( \tau) \sim \frac{\tau \psi (\tau)}{\langle \tau \rangle},
 \end{equation}
 where $\langle \tau \rangle$ is the mean flight duration \cite{AkimotoYamamoto2016a}.

\section{the double Laplace transform $\widetilde{P}(k,s)$ and the exact form of the second moment of $X(t)$ for $\nu>\gamma$}
Here, we represent the double Laplace transform $\widetilde{P}(k,s)$ as an infinite series expansion. 
Expanding $e^{-k\tau^\nu}$ in Eqs.~(\ref{psi_laplace_ta}) and (\ref{PSI_laplace_ta}), we have
\begin{equation}
\widetilde{{\color{black}\phi}}_{|v|}(k,s) \cong \widetilde{ \psi} (s) - \langle \tau^\nu \rangle k + O(k^2)
\end{equation}
and
\begin{equation}
\widetilde{{\color{black}\phi}}_{|v|}(k,s) \cong \widetilde{ \psi} (s) + \frac{c s^\gamma}{|\Gamma (-\gamma)|} \sum_{n=1}^\infty \frac{(-1)^n}{n!}  
\Gamma(n\nu -\gamma) \left(\frac{k}{s^\nu} \right)^n
\end{equation}
for $\nu < \gamma$ and $\gamma < \nu$, respectively, where $\langle \tau^\nu \rangle \equiv \int_0^\infty \tau^\nu \psi (\tau)d\tau$. 
Moreover, we have 
\begin{equation}
\widetilde{{\color{black}\Phi}}_{|v|}(k,s) \cong \frac{1-\widetilde{ \psi} (s)}{s} + \frac{c}{|\Gamma (-\gamma)|s^{1-\gamma}} \sum_{n=1}^\infty \frac{(-1)^n}{n!}  
\frac{ \Gamma(n\nu -\gamma + 1)}{\gamma + (1-\nu)n} \left(\frac{k}{s^\nu} \right)^n
\end{equation}
for $\gamma < 1$. Using Eq.~(\ref{montroll-weiss-like}), we have 
\begin{eqnarray}
\widetilde{P}_{|v|}(k,s) &=&  \frac{cs^\gamma}{s} \left[ 1 +  \frac{1}{|\Gamma (-\gamma)|} \sum_{n=1}^\infty 
\frac{1}{n!} \frac{ \Gamma(n\nu -\gamma +1)}{\gamma + (1-\nu)n} \left(-\frac{k}{s^\nu} \right)^n \right]
\frac{1}{ cs^\gamma  + \langle \tau^\nu \rangle k + O(k^2)} \nonumber\\
 &=&  \frac{1}{s}\left[ 1 + \sum_{n=1}^\infty 
\frac{1}{n!} \frac{\gamma \Gamma(n\nu -\gamma + 1)}{\Gamma (1-\gamma)\{\gamma + (1-\nu)n\}} \left(-\frac{k}{s^\nu} \right)^n \right]
\left[ 1 +   \frac{\langle \tau^\nu \rangle }{ c} \frac{k}{ s^\gamma} + O(k^2/s^\gamma)  \right]^{-1}
\label{montroll-weiss-like1}
\end{eqnarray}
for $\nu < \gamma$ and
\begin{eqnarray}
\widetilde{P}_{|v|}(k,s) &=&  \frac{cs^\gamma}{s} \left[ 1 +  \frac{1}{|\Gamma (-\gamma)|} \sum_{n=1}^\infty 
\frac{1}{n!} \frac{ \Gamma(n\nu -\gamma +1)}{\gamma + (1-\nu)n} \left(-\frac{k}{s^\nu} \right)^n \right]
\left[ cs^\gamma -  \frac{c s^\gamma}{|\Gamma(-\gamma)|} \sum_{n=1}^\infty \frac{  \Gamma(n\nu -\gamma)}{n!}
 \left(-\frac{k}{s^\nu} \right)^n \right]^{-1} \nonumber\\
 &=&  \frac{1}{s}\left[ 1 + \sum_{n=1}^\infty 
\frac{1}{n!} \frac{\gamma \Gamma(n\nu -\gamma + 1)}{\Gamma (1-\gamma)\{\gamma + (1-\nu)n\}} \left(-\frac{k}{s^\nu} \right)^n \right]
\left[ 1 -  \sum_{n=1}^\infty \frac{ \gamma \Gamma(n\nu -\gamma)}{n! \Gamma (1-\gamma)} \left(-\frac{k}{s^\nu} \right)^n \right]^{-1}\nonumber\\
 &=&  \frac{1}{s}\left[ 1 + \sum_{n=1}^\infty 
\frac{1}{n!} \frac{\gamma \Gamma(n\nu -\gamma + 1)}{\Gamma (1-\gamma)\{\gamma + (1-\nu)n\}} \left(-\frac{k}{s^\nu} \right)^n \right]
\left[ 1 +  \sum_{m=1}^\infty \left\{\sum_{n=1}^\infty \frac{ \gamma \Gamma(n\nu -\gamma)}{n! \Gamma (1-\gamma)}
 \left(-\frac{k}{s^\nu} \right)^n \right\}^m\right]
\label{montroll-weiss-like2}
\end{eqnarray}
for $\nu > \gamma$.

The coefficient of the term proportional to $k$ in Eq.~(\ref{montroll-weiss-like2}) is 
\begin{equation}
\frac{-1}{s^{1+\nu}} \left[ \frac{\gamma \Gamma(\nu-\gamma +1)}{(\gamma + 1 - \nu) \Gamma(1-\gamma)} + \frac{\gamma \Gamma(\nu - \gamma)}{\Gamma (1-\gamma)}\right].
\label{coeff-mean}
\end{equation}
Moreover, by considering the coefficient of the term proportional to $k^2$ in Eq.~(\ref{montroll-weiss-like2}), 
the leading term of the second moment of $X(t)$ in the Laplace space ($s\to 0$) can be represented as  
 \begin{equation}
\left. \frac{\partial^2 \widetilde{P}_{|v|}(k,s)}{\partial k^2}  \right|_{k=0} 
\sim \frac{M_2 (\nu, \gamma)}{s^{1+2\nu}},
\end{equation}
where 
\begin{equation}
 M_2( \nu, \gamma)= \frac{2\gamma \Gamma (2\nu -\gamma)}{(2-2\nu + \gamma)\Gamma (1-\gamma)}
+ \frac{2\gamma^2 \Gamma (\nu -\gamma)^2}{(1+\gamma -\nu) \Gamma(1-\gamma)^2}.
\label{M2_nu}
\end{equation}
It follows that the asymptotic behavior of $\langle X(t)^2 \rangle$ is given by Eq.~(\ref{moments_largernu}) 
with $n=2$.

{\color{black}Since $\langle X(t)^n \rangle$ grows as $\langle X(t)^n \rangle\propto t^{n\nu}$, one can define $M_n(\nu, \gamma)$  as}
\begin{equation}
\langle X(t)^n \rangle \sim \frac{\Gamma(1+\nu)^n M_n(\nu, \gamma)}{\Gamma (1 + n\nu) M_1(\nu, \gamma)^n} 
\langle X(t) \rangle^n.
\end{equation}
It follows that the random variable $X(t)/\langle X(t) \rangle$ converges in distribution to a random variable $M_{\nu,\gamma}$
which depends on both $\nu$  and $\gamma$. More precisely, one obtains 
\begin{equation}
\langle e^{-z M_{\nu,\gamma}} \rangle \sim \sum_{n=0}^\infty 
\frac{\Gamma(1+\nu)^n M_n(\nu, \gamma)}{n!\Gamma (1 + n\nu) M_1(\nu, \gamma)^n} 
(-z)^n.
\end{equation}
Therefore, the PDF of the normalized time average defined by $X(t)/t^\nu$ converges 
to a non-trivial distribution that is different from the Mittag-Leffler distribution (see Fig.~\ref{pdf-ta}).

\begin{figure}
\includegraphics[width=.9\linewidth, angle=0]{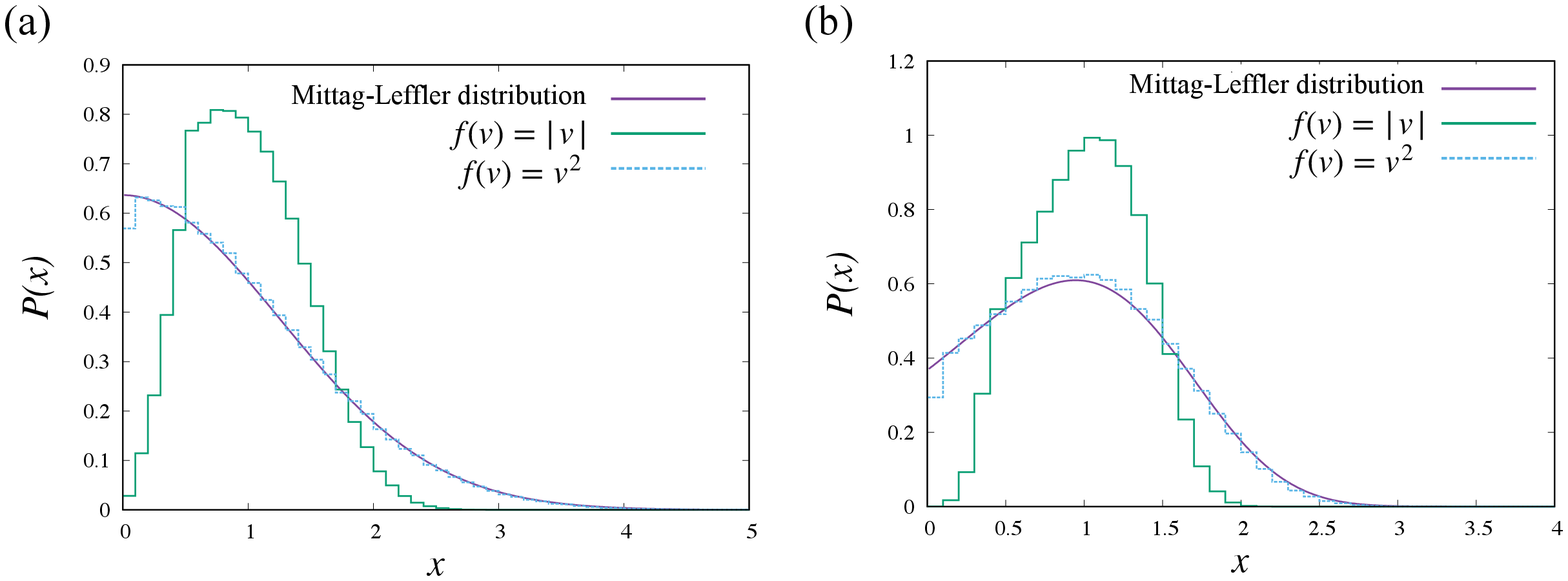}
\caption{Probability density function $P(x)$ of time averages for (a)  $\gamma=0.5$ and $\nu=0.7$ and (b)  $\gamma=0.7$ and $\nu=0.8$. 
The solid  and the dotted histograms represent PDFs obtained by numerical simulations 
for $f(v)=|v|$ and $f(v)=v^2$, respectively.  The solid line is the PDF of the Mittag-Leffler distribution.
Note that the PDFs for $f(v)=v^2$ follow the Mittag-Leffler distribution with order $\gamma=0.5$ and 0.7 in case (a) and (b), respectively. 
On the other hand,  the PDFs for $f(v)=|v|$ depend on the exponent $\gamma$ as  well as $\nu$, implying that the PDFs are different from
the Mittag-Leffler distribution. A similar distribution was found in distributional limit theorems of time-averaged observables in infinite ergodic theory \cite{Akimoto2015}.}
\label{pdf-ta}
\end{figure}

\end{widetext}

%

\end{document}
%